# Incentive Compatibility, Scalability and Privacy in real time Demand Response

G. Tsaousoglou, K. Steriotis, N. Efthymiopoulos, P. Makris, E. Varvarigos

*Abstract*--**The high penetration of Renewable Energy Sources in modern smart grids necessitated the development of Demand Response (DR) mechanisms as well as corresponding innovative services for the emerging flexibility markets. From a game-theoretic perspective, the basic key performance indicators (KPIs) for such DR mechanisms are: efficiency in terms of social welfare, practical applicability, and incentive guarantees, in the sense of making it a dominant strategy for each user to act truthfully according to his/her preferences, leaving no room for cheating. In this paper, we propose a DR architecture, including a mechanism based on Ausubel's clinching auction and a communication protocol, that provably guarantee both efficiency and truthful user participation. Practicality/easiness of participation is enhanced via simple queries, while user privacy issues are addressed via a distributed implementation. Simulation results confirm the desired properties, while also showing that the truthfulness property becomes even more important in markets where participants are not particularly flexible.**

## I. Introduction

Serving the energy demand in peak demand times might be quite expensive for the grid operator, because of the need to constantly maintain costly energy reserves. Also, in regions with high penetration of Renewable Energy Sources (RES), adjusting the demand to meet the intermittent generation can enhance the efficiency and economic viability of the system. As a result, the idea of offering monetary incentives (rewards) to consumers in order to decrease their consumption at peak demand times is getting a great deal of attention both from the research community and the Industry. Such techniques are generally referred to as Demand Response (DR). More specifically, when there is a need for reducing energy consumption in real-time, an ad-hoc market is created where the operator offers to buy consumption reduction from the users. Users participate in such a DR event by offering their consumption flexibility in exchange for monetary compensation.

In the modern smart grid, each user (consumer) has a smart meter that measures his/her consumption at all times. The grid operator can assess the aggregated consumption of users at a particular part of the grid in real-time. Users are interested in their own payoff, which results from the reward they receive and the discomfort they experience from reducing their energy consumption. On the other hand, the operator is interested in the reduction of the aggregated consumption at peak times. Assuming strategic user behavior, the above setting turns into a game, since each user's payoff is dependent on the actions of other users. In more detail, discomfort could be modeled through a local function, so that it is expressed in monetary terms. However, users are usually not capable of capturing their preferences in a closed form mathematical function and even if they were, they are reluctant to reveal their preferences. Rather, it is more natural for the users to simply take actions (e.g. turn appliances on/off, or adjust power consumption) in response to price signals.

An intermediate entity is assumed to resolve the formulated game and clear the ad-hoc flexibility market described above. We refer to this entity as the Flexibility Service Provider (FSP). The FSP is assumed to be an independent entity with the objective of coordinating the flexibility trading in the most efficient way. Formally, in economics, the "most efficient way" is characterized by the concept of maximizing the social welfare, defined as the aggregated payoff of all market participants. However, the users' local functions (related to their flexibility/comfort levels and consumption habits) are private to each user. This makes the task of the FSP quite challenging, especially when we consider users who act strategically and might misrepresent their local function if that makes them better-off.

In this paper, we propose a DR architecture through which FSPs will be able to optimally resolve the aforementioned game. In particular, we draw on concepts of mechanism design theory in order to define an iterative, auction-based mechanism, consisting of an *allocation rule* and a *payment rule*. The *allocation rule* refers to the way that the FSP decides upon how much consumption reduction will be allocated to each user according to the feedback obtained through the auction process. The *payment rule* refers to the way the FSP decides upon the reward of each user for his/her allocation, provided that the user makes the corresponding contribution. Through the auction procedure, the FSP exchanges messages with the users in the form of queries. A query in our case is a price signal communicated from the FSP to the user, to which the latter responds with his/her preferred action (i.e. consumption reduction) according to this signal. Note that a user may respond untruthfully if he/she finds that to be in his/her interest.

A mechanism is generally evaluated by: i) its performance in terms of social welfare, i.e. efficiency, ii) the tractability of

This work received funding from the European Union's Horizon 2020 research and innovation programme under grant agreement No. 731767 in the context of the SOCIALENERGY project.

G. Tsaousoglou, K. Steriotis, N. Efthymiopoulos, P. Makris are with the National Technical University of Athens (NTUA), Greece. E. Varvarigos is with Monash University, Melbourne, Australia.

the outcome, and iii) its incentive guarantees. The first two are commonly addressed in the literature and point to the allocation's efficiency and the mechanism's convergence time and consequent scalability. In contrast, the third requirement (that points to truthful participation) is widely overlooked in the DR literature. In the few cases where truthfulness is addressed, it comes with a sacrifice of practical implementation ability and user privacy. In the rest of this section we analyze what the third requirement is about and how it is handled in the state-of-the-art DR studies.

User strategies in games such as the one described above are subject to thorough study and discussion. Mechanism design theory classifies a mechanism's incentive guarantees with respect to how users are expected to act when participating in it. The strongest guarantee is called Dominant Strategy Incentive Compatibility (DSIC). We say that a mechanism is DSIC when it is at each user's best interest to truthfully implement his/her true preferences at any query, regardless of what other users do.

Surprisingly, the vast majority of studies in the DR literature do not provide any guarantees as we will analyze shortly. This drawback is typically rationalized by assuming that an individual user's load is very small compared to the whole system's aggregated load and thus the user can be approximated as a price taker (his/her actions, taken alone, have no effect on the system's dynamics). Under this assumption, each user implements his/her most favorable action (consumption decision), assuming the actions of other users to be constant. This process is repeated until an equilibrium is reached. The users are typically modeled to iteratively implement their best-response every time they are asked a query, i.e., they decide upon their preferred consumption upon receiving a price signal. This strategy updating procedure is called best-response dynamics. As analyzed in [1], such myopic "local rationality" does not necessarily imply "global rationality", i.e., given an iterative mechanism, it is not always to the user's best interest to repeatedly best-respond. Rather, a user might be better-off by submitting false bids through the process.

Best-response dynamics converges to an efficient allocation under the price-taking assumption described above. Nevertheless there are several use cases in which the assumption of price-taking behavior is rather strong and unjustified. For example, a large industrial consumer's actions may have a significant effect on the system. Also, when it comes to DR-events, the users called to participate are often required to be in a particular geographic location where congestion problems arise, in which case the relevant user population is not large. Another example includes islanded micro grids formed at neighborhood level, especially ones with high RES penetration. In such use cases, the number of users in the formulated game is drastically reduced. This means that a single user's actions may no longer be insignificant and a mechanism implemented in best-response strategies fails to capture user incentives. As a result, users are better expected to behave strategically, and strategic behavior may compromise the mechanism's efficiency [2]. In this paper we also address the third requirement, defined as the capability of the mechanism to provoke strategic users to act truthfully in accordance with their preferences, which is overlooked in most of the DR literature. Moreover, we do so via an indirect and practical mechanism, which allows for distributed and privacy-preserving implementation, in contrast to the few studies that consider incentive guarantees that do not exhibit these characteristics.

The rest of this paper is organized as follows. In Section 2, we present a literature review of DR studies from the perspective of incentive guarantees. In Section 3, we present the model assumed. In Section 4, we present the problem formulation. In Section 5, we present and analyze the proposed auction mechanism and prove that it has the desired properties. In Section 6, we demonstrate the performance and verify the properties of the proposed system. Finally, in Section 7 we describe a privacy-preserving communication protocol that can implement the proposed mechanism.

## II. RELATED WORK

In the DR architectures/frameworks that have appeared in the literature, the end user is typically modeled as a selfish player who participates in the mechanism with the purpose of maximizing his/her own payoff. The user's preferences are widely modeled as a convex function (e.g. [3]-[5]) in accordance with microeconomic theory [6]. However, studies differ on the way they model the behavior and the strategy of the users participating in the game. More specifically, there are three levels of behavior modeling, in increasing order of user rationality:

A) "naive", de-facto truthful users, assumed to always truthfully report their preferences
B) locally rational users, assumed to apply a myopic best-response process (maximizing their payoff at each iteration of the mechanism as if it were the last iteration)
C) strategic, globally rational users, who are aware of the mechanism's structure and apply a strategy that maximizes their final payoff (possibly by submitting false responses).

Several studies either assume naive users of category A ([7]-[15]) or assume no user preferences and perform central optimization for the scheduling problem (e.g. [16]-[17]).

The majority of DR works assume "price-taking users" which translates to category B, i.e., locally rational users. Static-pricing approaches (e.g. [18]-[19]), as well as typical dual decomposition approaches (including [3]-[5] and [20]-[25]), assume users of category B. Under the price-taking assumption, the solution concept is that of a competitive equilibrium. A market-clearing pricing approach brings the system to competitive equilibrium via an iterative best-response process, and the final allocation maximizes the social welfare. However, as described above, in many use cases (such as emerging local energy communities [26], [27] islanded micro-grids, etc) the price-taking assumption no longer holds and the efficiency of these mechanisms is compromised [2]. In mechanism design terms, the mechanisms of the first two categories are not *incentive compatible*, because a strategic user can benefit by manipulating his/her responses.

Few works consider user incentives. When considering strategic users (of category C), the mechanism designer is confronted with a trade-off: the Vickrey-Clarke-Groves (VCG) mechanism is the unique welfare maximizing mechanism implemented in dominant (and not best-response) strategies, meaning that either a VCG-based approach is taken [26]-[27] or welfare maximization is compromised [30]-[34].

The main problem with the VCG approaches [26]-[27] is that they require users to reveal their whole set of preferences to the FSP, while the latter makes all the calculations and decides the allocation and the rewards. This is clearly impractical, since real users generally can't express their preferences in closed-form mathematical functions and even when they can, they are not happy to compromise their privacy by sharing their whole set of preferences with the FSP. In this paper, we opt for a VCG-like approach, so as to achieve social welfare maximization, but we omit the direct-revelation approach of the typical VCG mechanism. Instead, we design an iterative auction mechanism based on Ausubel's clinching auction, in which users are only required to make decisions regarding their consumption in the presence of price signals. By adopting this approach, we implement the efficient VCG outcome but also allow for a distributed implementation and a privacy-preserving communication protocol.

Summarizing the above, our proposed DR architecture: i) is suitable for a distributed implementation (unlike [26]-[27]), ii) achieves the VCG outcome and does not sacrifice efficiency (unlike [30]-[31]), and iii) is incentive compatible (unlike studies that assume users of categories A and B).

## III. SYSTEM MODEL

We consider a flexibility market comprised of an FSP and a set $N \triangleq \{1,2,\dots,n\}$ of $n$ self-interested consumers, hereinafter referred to as users. We also consider a discrete representation of time, where continuous time is divided into timeslots $t \in T$ of equal duration $s$, where set $T \triangleq \{1,2,\dots,m\}$ represents the scheduling horizon. Each user possesses a number of controllable appliances, with each appliance bearing an energy demand. Since demands of different appliances are assumed independent and are not coupled, we can consider one appliance per user for ease of presentation and without loss of generality. We denote by the set of appliances.

### User & appliance modeling

An appliance requires an amount of energy for operation. For example, if an appliance's operating power is 1Watt, and $s = 1$ hour, then the energy that the appliance consumes in one timeslot of operation is $1Wh$. This energy consumption is measurable in real-time and can be shed if the user wishes. In particular, we consider controllable loads, meaning that the user can modify consumption upon request, in exchange for monetary compensation. Such a request for consumption modification is called a DR-event. Upon a DR-event asking for reduction of the real-time consumption in timeslot $t$, user $i$ can respond by reducing his/her consumption by a quantity $q_i^t$, assumed to be positive ($q_i^t \geq 0$), without loss of generality.

Also, $q_i^t$ is characterized by its feasible set $Q_i$ (defined by a set of constraints on $q_i^t$) and the discomfort function $d_i(q_i^t)$ of user $i$. The discomfort function is private to each user and expresses the minimum compensation in monetary units ($) that a user requires, in order to reduce his/her consumption by the corresponding amount. The discomfort as a function of $q_i^t$ can take various forms, depending on the appliance. We make the following assumptions on the form of function $d_i(q_i^t)$:

*Assumption 1*. Zero consumption reduction, brings zero discomfort to the user:
$$d_i(0) = 0$$

*Assumption 2*. The discomfort function is non-decreasing in $q_i^t$:
$$q_{iA}^t \geq q_{iB}^t \Leftrightarrow d_i(q_{iA}^t) \geq d_i(q_{iB}^t)$$
Assumption 2 says that consuming more does not make the user less comfortable.

*Assumption 3*: The discomfort function is upward sloping, meaning that additional increase of $q_i^t$ brings increasing discomfort to the user:
$$q_{iA}^t \geq q_{iB}^t \Leftrightarrow d_i(q_{iA}^t + \varepsilon) - d_i(q_{iB}^t + \varepsilon) \geq d_i(q_{iA}^t) - d_i(q_{iB}^t), \quad \forall \varepsilon, q_{iA}^t, q_{iB}^t > 0.$$

In order to incentivize users to reduce their consumption, the FSP offers a reward $r_i(q_i^t)$. A user's utility is defined as the difference between his/her discomfort for the consumption reduction realized and the reward he/she received for this reduction is
$$U_i = \sum_{t \in T}[r_i(q_i^t) - d_i(q_i^t)]. \quad (1)$$

In order to offer the rewards $r_i(q_i^t)$, the FSP draws on the reward offered by the operator who requests the reduction as described in the following subsection.

### DR-event and the FSP

Let $L^t$ denote the aggregated consumption of all users in $N$, as seen by the operator, within a certain time interval $t$. Upon a DR-event, the operator (e.g. the DSO that operates the smart grid) asks for a reduction of the users' aggregated consumption during a certain time interval and offers monetary incentives to the FSP towards its realization. Let $D^t$ denote the reduction in the aggregated consumption at $t$. The incentive (reward) is implemented as a per-unit compensation for the electricity units of reduced consumption. The cost of serving the aggregated energy consumption is typically modeled with quadratic functions [3]-[5] and [20]-[25] as explained in [35]. In this paper, we adopt the same approach and in direct analogy we assume that the compensation that is offered to the FSP by the operator, can be modeled as a concave function of $D^t$. For the purpose of being specific, we adopt here a polynomial function $R^t(D^t)$ of a specific form
$$R^t(D^t) = a \cdot D^t - b \cdot (D^t)^2, \quad D^t \in [0, L^t] \quad (2)$$
where $a, b$ are positive parameters with $a \geq 2bL^t$. The proposed DR architecture is open to any other choice of $R^t(D^t)$, provided it is a concave function. Thus, we assume that upon a DR-event, the operator offers a marginal per-unit reward
$$\lambda = \frac{d(R^t(D^t))}{d(D^t)} \quad (3)$$
for a consumption reduction of $D^t$ units.

The FSP is responsible for aggregating the users' participation in the DR-event, coordinating their actions, and dividing the compensation profits (rewards) among the users. We assume a communication network, built on top of the electricity grid, through which the FSP can monitor each user's consumption and exchange messages with the users.

## IV. PROBLEM FORMULATION

With respect to the system described above, we would like to facilitate the allocation of consumption reduction among the users so as to maximize social welfare. Social welfare is defined as the difference between the revenues $R^t(D^t)$ that the FSP receives from the operator for the consumption curtailment $D^t$ and the sum of the discomfort that this curtailment causes to its users. This problem can be formulated from Eqs. (4) and (4a) below:

$$\max_{q_i^t \in Q_i, i \in N} \{R^t(D^t) - \sum_{i \in N}[d_i(q_i^t)]\} \quad (4)$$
$$s.t. \; D^t = \sum_{i \in N} q_i^t \quad (4a)$$

The problem defined by Eqs. (4) and (4a) is a convex optimization problem and could be solved efficiently if the local functions $d_i(q_i^t)$ were known (or truthfully disclosed). However, $d_i(q_i^t)$ of each user is not known and thus, problem (4) is typically solved via dual decomposition in the DR literature (see [3]-[5] and [20]-[25]). This approach, however, is not incentive compatible as we will analyze shortly. In particular, the final allocation of the dual decomposition approach is identical to that obtained through the ascending English auction (see algorithm 2 of [3]), which halts when supply equals demand. More specifically, in the system model described and in case of an English auction, the FSP would iteratively increase a per-unit reward $\lambda$ asking the users their consumption reduction $q_i^t(\lambda)$ at each per-unit reward $\lambda$ (auction query). At each iteration, each user $i$ responds with his/her preferred $q_i^t(\lambda)$. A truthful (locally optimal) response by user $i$, denoted as $\tilde{q}_i^t(\lambda)$, is one that maximizes $i$'s utility for reward $\lambda$. This is mathematically formulated as the solution to maximization problem (5):

$$\tilde{q}_i^t(\lambda) = \text{argmax}_{q_i^t \in Q_i, i \in N} \{\lambda \cdot q_i^t - d_i(q_i^t)\} \quad (5)$$

Clearly, $\tilde{q}_i^t(\lambda)$ is non-decreasing in $\lambda$, since $q_i^t \geq 0$. The auction terminates when $\lambda$ reaches a value for which $\sum_{i \in N} q_i^t(\lambda) = D^t(\lambda)$. The final price is commonly called the market-clearing price and it is denoted here as $\lambda_{mc}$. The allocation at $\lambda_{mc}$ is efficient if the users truthfully report their $q_i^t$ at each FSP query. However, truthful report may not be the best strategy for every user. To illustrate this, we present the following example:

*Illustrative example*

Consider two users and a given timeslot $t$. User 1 operates a load with power consumption 10 kW while user 2 operates a 50 kW load. Now suppose they participate in a DR event and their discomfort function is $d_i(q_i^t) = \omega_i \cdot (q_i^t)^2$, $i \in \{1,2\}$, where their true flexibility parameters are $\omega_1 = \omega_2 = 0.1$. The reward function is $R^t(\Delta L^t) = 5 \cdot (\Delta L^t)$. Should they act according to their true discomfort function parameters, their utilities (given from Eq. (1)) at equilibrium would be $U_1 = U_2 = 4.875$ units. In case User 2 acts untruthfully according to $\omega_2^{fake} = 0.2$, his utility at equilibrium will be $U_2 = 7$. Therefore, the best strategy of User 2 is to be untruthful. ∎

The previous *example* demonstrates how the market-clearing approach builds on the assumption that users behave myopically, by truthfully maximizing their utility at each iteration. However, a DR-event will involve smart players (e.g. industrial consumers, aggregators) and it will not take long before users realize that they can benefit from engineering untruthful responses. The problem is that if we relax the truthfulness assumption and consider strategic users, market-clearing methods (e.g., the English auction presented above) no longer result in efficient allocations. For this reason it is very important to design a mechanism that is not only efficient but also incentive compatible.

In order to facilitate the description of the proposed mechanism, we first present the Vickrey-Clarke-Groves (VCG) mechanism, which is the unique mechanism that makes it a dominant strategy (DSIC as analyzed in the introduction) for each user, to act truthfully, i.e. in accordance with his/her real discomfort function [36]. Let $N_{-i}$, denote the set of users, excluding user $i$. The VCG payment rule is the so called "Clarke pivot rule", which calculates a reward $r_i$ equal to $i$'s "externality". In other words, it rewards each user $i$ with an amount equal to the difference that $i$'s presence makes in the social welfare of other users $j \in N_{-i}$:

$$r_i(q_i^t) = R^t(\sum_{j \in N_{-i}} q_j^t) - \sum_{j \in N_{-i}} d_j(q_j^t)$$
$$-R^t(\sum_{j \in N_{-i}} \hat{q}_j^t) + \sum_{j \in N_{-i}} d_j(\hat{q}_j^t), \quad (6)$$

where $q_j^t$ denotes the vector allocated to user $j$ when problem (4) is solved with user $i$ included in the system, and $\hat{q}_j^t$ denotes the vector allocated to user $j$ when the same problem is solved without user $i$'s participation.

In the direct VCG mechanism, users are asked to declare their local functions $d_i(q_i^t)$ to the FSP. Because of the Clarke pivot rule, it is a dominant strategy for each user to make a truthful declaration [36]. Thus, the efficient allocation that corresponds to the social welfare maximization problem can be calculated at the FSP side. In order to calculate the VCG rewards from Eq. (6), problem (4) is solved $|N| + 1$ times (one time with each user in $N$ absent to calculate the payments, plus one time with all users present to calculate the allocation). The major drawback of the direct VCG mechanism is the requirement that the users disclose their discomfort functions $d_i(q_i^t)$ to the FSP. This raises important issues such as a) Lack of privacy in case where users are reluctant to reveal local information (their discomfort function) and b) Difficulty in implementation in cases where users are unable to express their preferences (i.e., their discomfort function) in a closed form function.

In the next section, we propose a modification of Ausubel's Clinching auction [37], which allows for a distributed implementation of VCG as described in section VII, designed to tackle these issues. In particular, we opt for an iterative auction that:
i) facilitates user bids via auction queries, thus making the proposed architecture more easily implementable in practice
ii) engages users in the market and allocates consumption reduction gradually along the way, so that price discovery is facilitated on the users' side
iii) protects user's privacy via a properly designed communication protocol.

## V. Ausubel's Clinching Auction for DR-event participation

The Clinching Auction (CA) is a well-known ascending price auction (similar in fashion to English Auction) that halts when demand equals supply. However, in contrast to most auctions (including the English auction), allocation and rewards are not cleared exclusively at the final iteration. Rather, the goods (consumption reduction in our context) are progressively allocated as the auction proceeds and payments are also progressively built, while the auction design guarantees that the final allocation and payments coincide with the ones obtained through VCG. Thus, both allocation efficiency and incentive compatibility are achieved, while the aforementioned privacy and implementation drawbacks of the direct-VCG mechanism are effectively addressed.

In order for the CA to work in our setting, we need to reverse the price trajectory. In the proposed Modified Clinching Auction (MCA), the FSP begins with a per-unit reward $\lambda = \lambda_{max}$ which gradually decreases at each iteration. By Eq. (3), reward $\lambda_{max}$ is $\frac{dR^t(0)}{d\Delta L^t} = a$, which, as analyzed in Section 3, is the highest value possible given that $R^t$ is concave. Users respond by bidding their preferred consumption reduction $\tilde{q}_i^t(\lambda)$ for each $\lambda$. We represent the user's response at $\lambda$ as the solution to the user utility maximization problem (which is formally defined in Eq. (5) of the previous section).

The user's objective function is concave in $q_i^t$, since $\lambda \cdot q_i^t$ is linearly increasing and $d_i(q_i^t)$ is convex by *Assumption 3*. Also, the solution $\tilde{q}_i^t$ is increasing in $\lambda$, which means that the user's response $\tilde{q}_i^t$ gradually decreases as $\lambda$ decreases. Note that in the extreme and trivial case where $\lambda_{max} \cdot \sum_{i \in N}(\tilde{q}_i^t(\lambda_{max})) \leq R^t(D^t)$ the users would shut down everything and proportionally share the reward $R^t(D^t)$.

In MCA, the initial price is $\lambda_{max}$ and in each iteration $k$ the price $\lambda^k$ is reduced by a small positive number $\varepsilon$. The size of $\varepsilon$ adjusts the discretization level of MCA. For the decreasing reward auction that we propose, we relax constraint (4b) to the inequality

$$D^t \geq \sum_{i \in N} q_i^t. \quad (7)$$

Consider an arbitrary iteration $k$ of the MCA and let $D^t(\lambda^k)$ denote the operator's desired reduction for per-unit reward $\lambda^k$. The central idea of the MCA is the following: if there is a set $N^{\not{j}} \subset N$ for which we have

$$D^t(\lambda^k) - \sum_{j \in N^{\not{j}}}\left(\tilde{q}_j^t(\lambda^k)\right) > 0 \quad (8)$$

then we allocate a reduction equal to $\zeta_i^k = D^t(\lambda^k) - \sum_{j \in N^{\not{j}}}\left(\tilde{q}_j^t(\lambda^k)\right)$ to each user $i \notin N^{\not{j}}$ at a per-unit reward $\lambda^k$. We then say that user $i$ "clinched" $\zeta_i^k$ units. The MCA auction terminates when set $N^{\not{j}}$ that satisfies condition (8) and set $N$, are equal, that is, constraint (7) is satisfied. After that, it allocates the remaining $D^t(\lambda^{k-1})$ proportionally to the users that bid in the second-to-last iteration.

The critical advantage of the Clinching auction is that it allocates different amounts of units at different rewards, and the units that a user clinches do not depend on his/her own bid but only on the other users' bids. The algorithm that implements MCA is presented in Table 1.

**Table 1. The MCA algorithm**

1. Initialize $\lambda^0 = \lambda_{max}, q_i^t(\lambda_{max}), D^t(\lambda_{max}), k = 0$
2. **while** $D^t(\lambda^k) < \sum_{i \in N}(\tilde{q}_i^t(\lambda^k))$
3.     **if** there exists $N^{\not{j}}: \sum_{j \in N^{\not{j}}}\left(\tilde{q}_j^t(\lambda^k)\right) < D^t(\lambda^k)$
4.        clinch units $\zeta_i^k = D^t(\lambda^k) - \sum_{j \in N^{\not{j}}}\left(\tilde{q}_j^t(\lambda^k)\right)$
for all $i \notin N^{\not{j}}$ at per-unit reward $\lambda^k$
5.     **else**
6.        set $\lambda^{k+1} = \lambda^k - \varepsilon$ and $k = k + 1$
7.        ask each user a reduction query for $\lambda^k$ and collect the responses $q_i^t(\lambda^k)$
8.        ask the operator for the desired total reduction $D^t(\lambda^k)$ at per-unit-reward $\lambda^k$
9. **End while**
10. Clinch units
$$\zeta_i^k = \left(q_i^t(\lambda^{k-1}) - \sum_{\eta=0}^{k-1}\zeta_i^\eta\right) \cdot \frac{D^t(\lambda^{k-1})}{\sum_{i \in N} q_i^t(\lambda^{k-1})}$$
at per-unit reward $(\lambda^{k-1})$, for each $i \in N$

We are now in a position to prove the optimality of MCA in terms of social welfare performance:

*Theorem 1:* The social welfare loss at the final allocation of MCA is within $(\varepsilon^2 + \lambda_{max} \cdot \varepsilon)/2b$ of the maximum possible.

*Proof:* The value of $\lambda$ at which $D^t = \sum_{i \in N}(\tilde{q}_i^t)$ is defined as $\lambda_{mc}$, which gives

$$D^t(\lambda_{mc}) = \sum_{i \in N}\left(\tilde{q}_i^t(\lambda_{mc})\right). \quad (9)$$

Let $\hbar$ denote the number of iterations until the auction halts, that is,

$$\hbar = \left\lceil \frac{\lambda_{max}-\lambda_{mc}}{\varepsilon}\right\rceil, \quad (10)$$

where $\lceil\cdot\rceil$, denotes the rounding to the nearest integer above. We have

$$\left\lceil \frac{\lambda_{max}-\lambda_{mc}}{\varepsilon}\right\rceil \leq \hbar \leq 1 + \left\lceil \frac{\lambda_{max}-\lambda_{mc}}{\varepsilon}\right\rceil \quad (11)$$

After the last clinchings (line 10 of the algorithm) we have efficiently allocated $D^t(\lambda^{\hbar-1})$ reduction units to the users. The remaining $D^t(\lambda_{mc}) - D^t(\lambda^{\hbar-1})$ are not allocated and this causes the loss of welfare ($W_{loss}$) that is depicted as the grey area in Fig.1, where the red line represents $D^t(\lambda)$ and the blue line represents $\sum_{i \in N} \tilde{q}_i^t(\lambda)$.

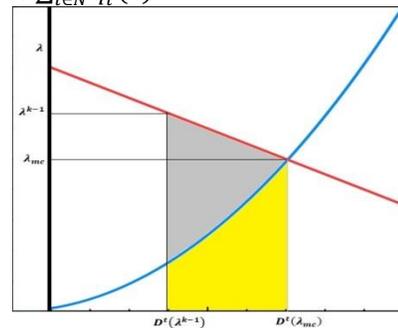

Fig. 1 $D^t(\lambda)$ and $\sum_{i \in N}(\tilde{q}_i^t(\lambda^k))$ as a function of $\lambda$

Since we remain agnostic of the closed form of $\sum_{i \in N}(\tilde{q}_i^t(\lambda^k))$, we assume the worst case and calculate an upper bound on the sum of the grey plus the yellow area of Fig 1:

$$W_{loss} \leq \lambda_{mc}\left(D^t(\lambda_{mc}) - D^t(\lambda^{\hbar-1})\right) + \frac{1}{2}(\lambda^{\hbar-1} - \lambda_{mc})\left(D^t(\lambda_{mc}) - D^t(\lambda^{\hbar-1})\right).$$

By substituting $D^t(\lambda) = \frac{a-\lambda}{2b}$ from Eq. (3), we get

$$W_{loss} \leq \frac{\lambda_{mc}(\lambda^{k-1}-\lambda_{mc})}{4b} + \frac{\lambda^{k-1}(\lambda^{k-1}-\lambda_{mc})}{4b}$$
$$\leq \frac{(\lambda^{k-1})^2 - (\lambda_{mc})^2}{4b}.$$

By further substituting $\lambda^{k-1} = \lambda_{max} - \varepsilon(k-1)$ and also substituting $k$ from inequalities (11), using the left inequality when $k$ appears with a minus sign and the right inequality when it appears with a plus sign, we finally obtain
$$W_{loss} \leq \frac{\varepsilon^2 + \lambda_{max} \cdot \varepsilon}{2b},$$
completing the proof. ∎

In practice, for the relevant use cases of price-anticipating users (described in the introduction), the computational complexity of the MCA is small, which allows for a very small choice of $\varepsilon$. To emphasize this, it is useful to state the following corollary to Theorem 1:

*Corollary 1:* for $\varepsilon \ll 1$ the welfare loss grows linearly with $\varepsilon$.

Because the MCA includes a price-sensitive response also at the operator's side, we have to verify that the properties of efficiency and incentive compatibility still hold. This is proved in the following Propositions.

*Proposition 1:* Truthful bidding is a dominant strategy in MCA.

*Proof*: Fix an iteration $k$ and suppose that $i$ bids $q_{i,false}^t(\lambda^k) \neq \tilde{q}_i^t(\lambda^k)$ in that iteration. From step 4 of MCA, we see that $\zeta_i^k$ does not depend on $q_i^t$ but only on the other users' bids $q_j^t, j \neq i$. Thus, user $i$'s bid can affect $i$'s allocation only by changing the $\lambda$ at which the termination condition holds. This means that a false bid $q_{i,false}^t(\lambda^k)$ will make a difference to $i$, only if $k$ is the last iteration. However, by definition of $\tilde{q}_i^t(\lambda^k)$ (see Eq. (5)), any bid $q_{i,false}^t(\lambda^k) \neq \tilde{q}_i^t(\lambda^k)$ brings strictly lower utility to user $i$ at any iteration $k$. Thus, truthful bidding brings the highest utility to user $i$. ∎

Furthermore, the following properties of the VCG mechanism hold also for the MCA:

*Proposition 2:* MCA is *individually rational*, weakly *budget-balanced*, and achieves the maximum revenue for the FSP among all efficient mechanisms.

*Proof*: The MCA auction is welfare maximizing (by Theorem 1, for $\varepsilon$ small enough) and DSIC (by Proposition 1). However, the class of VCG mechanisms is the unique class that simultaneously achieves these two properties [38]. Thus, MCA terminates with the VCG allocation and payments, and it inherits the property of *individual rationality*. For the weak budget balance property, it suffices to show that our setting exhibits the no single-agent effect [38]. An environment exhibits no single-agent effect if the aggregated utility of $n-1$ users doesn't improve by adding a $n^{th}$ user to the system. This property holds in single-sided auctions with monotonous preferences [38], since dropping a user only reduces the competition for the remaining users, thus making them better-off.

Moreover by [36], the VCG mechanism maximizes the auctioneer's utility, which means that the FSP buys flexibility units from the users at the lowest possible price (among all efficient and individually rational mechanisms). ∎

## VI. PERFORMANCE DEMONSTRATION

In this section, we use simulations to demonstrate the advantages of the MCA and verify its properties. As a benchmark for comparison, we use the typical market-clearing pricing where all users receive a per-unit reward of $\lambda_{mc}$. Over a time horizon of 24 timeslots, we simulated two DR events, in timeslots 11 and 17 where there was a peak in the aggregated consumption. Parameters $a$ and $b$ of the reward function were set to $a = 3$ and $b = 0.02$ for both timeslots.

We used a simple model for the user's discomfort function:
$$d_i(q_i^t) = \omega_i^t \cdot (q_i^t)^2,$$
where parameter $\omega_i^t$ expresses the user's inelasticity in timeslot $t$. In order to obtain results for a wide range of parameters $\omega_i^t$, we pick $\omega_i^t$ from a random uniform distribution in $[0.5 \cdot \omega_f, 1.5 \cdot \omega_f]$ for $t = 11$ and in $[0.05 \cdot \omega_f, 1.5 \cdot \omega_f]$ for $t = 17$, where parameter $\omega_f$ will vary in our experiments. We set the step $\varepsilon = 10^{-5}$ in the MCA algorithm (Table 1). Figure 2 depicts the aggregated consumption along all 24 timeslots for $\omega_f = 1$, which shows the reductions in consumption corresponding to the DR events.

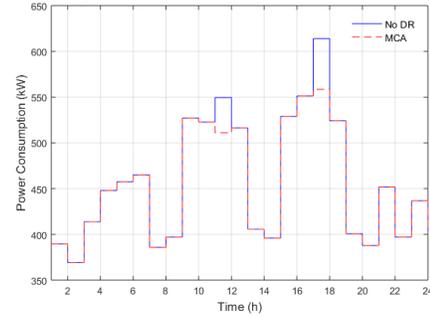

**Fig 2. Aggregated consumption as a function of time with and without DR events in timeslots 11 and 17**

In order to verify the truthfulness property and that a user can only lose by not being truthful, we assume that one user acts untruthfully by manipulating his/her $\omega_i$ for timeslot 17, while all other users act truthfully. The untruthful user is indexed by $ch$ (for cheater). The cheater's utility $U_{ch}$ is maximized for a certain choice of $\omega_{ch}$, denoted as $\omega_{ch}^{fake,*}$. Figure 3 shows $U_{ch}$ as a function of $\omega_{ch}$ (for $\omega_f = 5$).

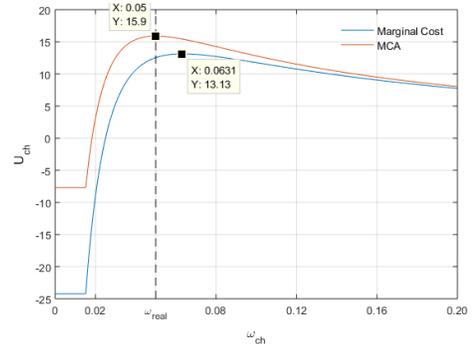

**Fig 3. Focal user's utility as a function of his/her choice of $\omega$**

The black vertical line represents the focal user's actual (real) $\omega$, denoted as $\omega_{real}$. For the MCA, the user's optimal choice of $\omega$ coincides with his/her real $\omega$, that is $\omega_{ch}^{fake,*} = \omega_{real}$, thus verifying Proposition 1.

Next, we investigated the effect that cheating has on the FSP's profits, denoted by $\Pi^{truthful}$ for the case where users act truthfully and by $\Pi^{cheat}$ for the case where they act according to what brings them the highest utility. Figure 4 shows that the ratio $\Pi^{cheat}/\Pi^{truthful}$ is maximized and is equal to 1 for the MCA, verifying our theoretical results. We also observe that the FSP's profit loss due to untruthfulness rises with $\omega_f$ (i.e. when users are less elastic), indicating that our scheme's truthfulness property becomes more important in markets where participants are not particularly flexible.

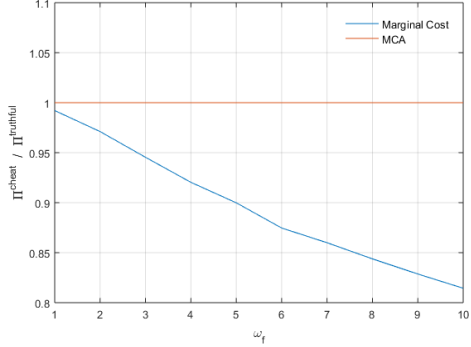

Fig 4. Ratio $\frac{\Pi^{cheat}}{\Pi^{truthful}}$ as a function of $\omega_f$

Finally, we simulated the DR-event for timeslot 17 for different values of $\varepsilon$, measuring the proportional welfare loss

$$W_{loss} = \frac{W_{opt} - W_{MCA}}{W_{opt}},$$

where $W_{opt}$ is the optimal welfare and $W_{MCA}$ is the welfare achieved by the MCA. The simulation results in Fig. 5 verify Corollary 1, which states that for small values of $\varepsilon$ the upper bound on the welfare loss grows linearly with $\varepsilon$.

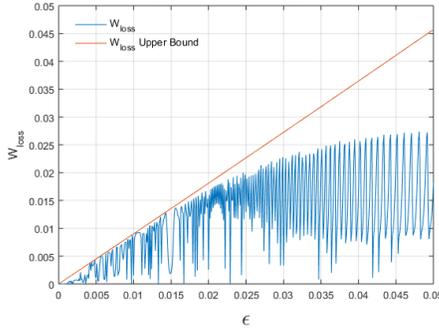

Fig 5. Proportional welfare loss of MCA as a function of the price step $\varepsilon$

## VII. PRIVACY-PRESERVING DISTRIBUTED IMPLEMENTATION

A major drawback of the direct VCG mechanism is that it requires each user to know and disclose his/her discomfort function to a central entity, e.g., the FSP. The MCA auction implements the VCG allocation and payments via an indirect mechanism. In this way users are only required to respond to FSP queries, instead of being required to communicate their discomfort function. This allows a distributed implementation of an efficient and truthful DR architecture. In what follows we present a distributed communication protocol that preserves privacy while simultaneously ensuring an efficient allocation.

The proposed DR architecture exploits [39] in order to execute MCA in a distributed fashion. In this way, the FSP does not have to learn the answers to the queries, which are instead acquired only by users in $N$ in a distributed fashion. Thus, the proposed DR architecture acts as a substrate that offers a service over which participating users cooperate in order to protect their personal data (i.e. their discomfort functions $d_i(\cdot)$) from the FSP. In order to achieve this, [39] uses the scheme proposed by Kademlia [40] in which each node (i.e., end user/energy consumer) is identified by a number (nodeID) in a specific virtual space. The nodeIDs do not serve only as identification, but they are also used by the Kademlia algorithm to store and locate values/data hashes (i.e., the answers to the FSP queries). This process is realized through a peer to peer routing service (implemented in the network application layer) that Kademlia offers. Towards this end, participating nodes create and dynamically maintain routing tables in a bottom up organized way. In fact, the nodeID provides a direct map to these data hashes by storing information on where to obtain them. The proposed algorithm is executed in three steps:

1. *Data insertion*: At each iteration $k$ of the algorithm, each user (node) $i$ stores its bid $\widetilde{q}_i^t(\lambda_k)$ in another random node $w$ through the use of the aforementioned [40] system. It is highlighted that $w$ is different for each $i$ and $k$ (as it is derived from the output of the hash function that Kademlia uses), and in this way collusion of two users (which is a requirement that [41] sets), or even collusion of a relatively small number of users to acquire data, will fail.

2. *Calculation of $\zeta_i^k(\lambda_k)$*: Kademlia organizes the participating nodes in a tree like structure. The proposed system exploits this structure in order to calculate the sum $\sum_{i \in N} \widetilde{q}_i^t(\lambda^k)$. To do so in a distributed way, node $j$ waits until all nodes with lower nodeID from it, inform $j$ on possible data values they have to send to $j$. This process continues recursively until the node with the highest id acquires the desirable data and then it calculates the sum. At this point, this node also receives $D^t(\lambda^k)$ from the FSP and checks the termination condition. If it doesn't hold, the node proceeds by broadcasting $\sum_{i \in N} \widetilde{q}_i^t(\lambda^k)$ and $D^t(\lambda^k)$ to all nodes through the use of Kademlia tree [40]. Thus, each node $j$ calculates $\zeta_i^k(\lambda^k)$ by subtracting the $\widetilde{q}_i^t(\lambda^k)$ value that is stored in it (which is not its own $\widetilde{q}_j^t(\lambda^k)$ value, and it doesn't know whose it is).

3. *Final allocation and payments calculation*: at the next iteration $k+1$, a different instance of Kademlia tree is created, so that $\zeta_i^{k+1}(\lambda^{k+1})$ is stored at a new node $g$, other than $j$. Thus, even in the case that a node is malicious, data privacy is not compromised. The tuple $A_i = \{\sum_{m=1}^{k} \zeta_i^m(\lambda^m), \sum_{m=1}^{k}[\zeta_i^m(\lambda^m) \cdot \lambda^m]\}$, which contains the allocation and payments of user $i$ up until iteration $k$, is passed from user $j$ to $g$. At the final iteration, the tuples $A_i$ are communicated to the FSP. Note that the FSP receives only the final allocation and payments for each user, i.e., only the sum of $\zeta_i^k(\lambda^k)$ and not all the intermediate values $\zeta_i^m(\lambda^m)$. This means that the FSP (and any other node for that matter) does not have the data to construct the entire local discomfort function $d_i(\cdot)$ of user $i$.

Note that the analysis above assumes that the service provider is honest-but-curious. By this we mean that the FSP is curious to know the discomfort functions of end users, but is also honest and will never attack the system in order to acquire them. In case of malicious FSP (i.e. with no hesitations to break the law), more strict privacy assumptions are needed, but this case is outside the scope of the present work.